\documentclass[aps,prd,twocolumn,superscriptaddress,showkeys]{revtex4-2}
\usepackage{epsfig}
\usepackage{graphicx}
\usepackage{amsmath,amssymb}

\newcommand{\bastar}{\begin{eqnarray*}}
\newcommand{\eastar}{\end{eqnarray*}}
\newskip\humongous \humongous=0pt plus 1000pt minus 1000pt

\newif\ifdtup

\relax
\newcommand{\bea}{\begin{eqnarray}}
\newcommand{\eea}{\end{eqnarray}}
\newcommand{\X}{\vec X}

\newcommand{\pd}{\partial}
\newcommand{\n}{\hat n}

\newcommand{\mn}{{\mu\nu}}

\newcommand{\F}{\vec F}

\newcommand{\A}{\vec A}
\newcommand{\hA}{\hat A}

\begin{document}
\title{Two Types of Gluons in QCD: Re-interpretation 
of ALEPH and CMS Gluon Jet Data}
\bigskip
\author{Y. M. Cho}
\email{ymcho0416@gmail.com}
\affiliation{School of Physics and Astronomy,
Seoul National University, Seoul 08826, Korea}
\affiliation{Center for Quantum Spacetime, 
Sogang University, Seoul 04107, Korea}  

\begin{abstract}
The Abelian decomposition of QCD tells that there are 
two types of gluons in QCD, the color neutral nurons 
and colored chromons, which have different color quantum numbers and thus behave differently. This strongly implies that QCD has two types of gluon jets, the nuron jet and chromon jet. One quarter of the gluon jets is made of 
the nuron jets which have sharper jet radius and smaller particle multiplicity, while three quarters of them are made of chromon jets which have broader jet radius and larger particle multiplicity. Moreover, the nuron jet 
has a distinct color flow which forms an ideal color 
dipole pattern, while the chromon jets have distorted dipole pattern. In this paper we provide five circumferential evidences of the existence of two types 
of gluon jets from the existing ALEPH and DELPHI data 
on $e \bar e \rightarrow Z \rightarrow b \bar b g$ 
three jet events and the CMS data on Pb-Pb heavy ion collision. 
\end{abstract}
\pacs{12.38.-t, 12.38.Aw, 11.15.-q, 11.15.Tk}
\keywords{Abelian decomposition of QCD, restricted potential, valence potential, monopole potential, two types of gluons, nuron, chromon, nuron jet, chromon jet, anormaly in particle multiplicity of gluon jet, anormaly in energy fragmantation of gluon jets, correlation between the low particle multiplicity and large energy fragmantation in gluon jets, anormaly in the gluon quenching in heavy ion collision}
\maketitle

\section{Introduction}

With the asymptotic freedom and the subsequent experimental confirmation of the gluon jet, the quantum chromodynamics (QCD) has become widely accepted as the theory of the strong interaction \cite{wil,ellis,gjet}. But it has many unresolved problems. The color confinement problem known as 
the millennium problem is the most outstanding one. 
Two popular conjectures to resolve this problem are 
the monopole condensation \cite{nambu,prd80,prl81} 
and Abelian dominance \cite{prd80,prl81,thooft,prd00}. 
The monopole condensation is based on the assumtion that, 
just as the Cooper pair (the electron pair condensation) confines the magnetic flux in superconductors, the monopole condensation (more precisely the monopole-antimonopole pair condensation) in QCD could confine the color electric flux. 
To prove this, however, we have to separate the monopole potential from the QCD gauge potential. Can we 
do this? 

On the other hand the Abelian dominance assumes that 
only the Abelian part of the gluon potential can make 
the confinement. This is because the non-Abelian part 
must be confined since they are colored, so that they 
can not play the role of the confiner. This is just like 
the prisoner who should be jailed can not play the role of 
the jailer. To prove this, however, we should be able to separate the Abelian part of the QCD potential gauge independently. How can we do this?      

There are other problems. Consider the proton made of three quarks. Obviously we need the gluons to bind the quarks 
in the proton. However, the quark model tells that 
the proton has no constituent gluon. If so, what is 
the difference between the binding gluon and constituent gluon? And how can we separate them gauge independently? 

Moreover, the simple group theory tells that two of 
the eight gluons corresponding to the diagonal generators of the color gauge group must be color neutral, while the other six which correspond to the off-diagonal generators must 
carry the color. This strongly implies that there are two types of gluons, the color neutral ones and colored ones, 
which should behave differently. Can we distinguish them?  

The bottom line here is to understand how different 
the non-Abelian dynamics of QCD is from the Abelian QED. Obviously all non-Abelian gauge groups have the maximal Abelian subgroups. Can we separate the Abelian subdynamics corresponding to the Maximal Abelian subgroup from 
the full non-Abelian dynamics? And how different is this Abelian subdynamics from QED? To answer this we should first be able to separate the Abelian subdynamics of QCD gauge independently. How can we do this? 
 
The Abelian decomposition of QCD could resolve these 
problems nicely \cite{prd80,prl81,prd01}. It decomposes 
the QCD gauge potential to two partsgauge independently, 
the restricted (Abelian) potential which nevertheless 
retains the full non-Abelian gauge degrees of freedom 
and the valence potential which transforms gauge covariantly. Moreover, it tells that the restricted potential is made of two parts, the non-topological Maxwellian Abelian potential and the topological Diracian monopole potential.  

The Abelian decomposition does not change QCD but 
allows us to understand the color dynamics more 
clearly, by revealing the important hidden structures 
of the conventional QCD. It tells that QCD has 
the restricted QCD (RCD) made of the restricted 
potential which describes the Abelian subdynamics 
of QCD which nevertheless retains the full color gauge degrees of freedom. This tells that the Abelian 
subdynamics of QCD is different from QED. RCD is made 
of not only the non-topological Maxwellian potential 
but also the topological Diracian monopole potential.

Moreover, it allows us to interprete QCD as RCD which 
has the valence potential as the colored source.  
This has deep implications. First, this decomposes 
the QCD Feynman diagrams in such a way that the color conservation is explicit, decomposing the gluon propagators to the nuron and chrom propagators. Second, this implies that the nuron (like the photon in QED) plays the role 
of the binding gluon and the chromon (like the quark) 
plays the role of the constituent gluon.

Third, this tells that what makes QCD different from 
QED is the topological Diracian monopole potential, 
which strongly implies that the monopole is responsible 
for the color confinement. Finally, this puts QCD to 
the background field formalism which provides us an ideal platform to calculate the QCD effective potential and 
porve the monopole condensation \cite{prd01,prd13,epjc19}. This is because we can treat the restricted part as 
the classical background and the valence part as 
the fluctuating quantum field. 

\begin{figure}
\includegraphics[height=2.5cm, width=7cm]{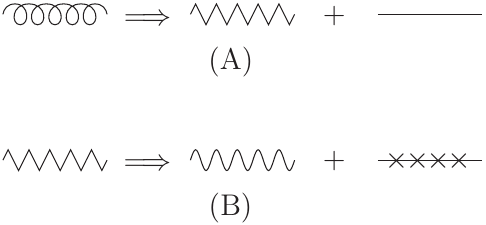}
\caption{\label{cdec} The Abelian decomposition of 
the QCD gauge potential. In (A) it is decomposed to 
the restricted potential (kinked line) and the chromon
(straight line) potential. In (B) the restricted potential 
is further decomposed to the nuron (wiggly line) and 
the monopole (spiked line) potentials.}
\end{figure}

This means that, after the confinement sets in,  
the perturbative QCD is made of the non-topological 
Maxwellian Abelian potential which describes the color 
neutral gluon (the nuron) and the gauge covariant 
valence potential which describes the colored gluon 
(the chromon). In other words, there are two types 
of gluons in QCD, the color neutral nuron and colored 
chromon. 

The Abelian decomposition is not just a theoretical proposition. The existence of two types of gluons can be verified by experiment \cite{uni19,prd23}. {\it The purpose 
of this paper is to provide five circumferential evidences 
of the existence of two types of gluons, re-interpreting 
the existing ALEPH and DELPHI gluon jet data coming from 
the Z boson decay and CMS gluon jet data coming from heavy 
ion collision. We argue that the known gluon jet anormalies and puzzles in the ALEPH, DELPHI, and CMS gluon jet data 
are in fact the evidence of the existence of two types of gluons in QCD.} 

\section{Abelian decomposotion of QCD: A Brief Review}

Consider the SU(2) QCD for simplicity
\begin{gather}
{\cal L} = -\frac14 \F_\mn^2.
\end{gather}
Choosing an arbitrary direction $\n$ as the Abelian direction and selecting the potential which parallelizes the Abelian direction imposing 
the condition 
\begin{gather}
D_\mu \n=0,	
\end{gather}
we can project out the restricted potential $\hA_\mu$ 
which describes the Abelian subdynamics of QCD from 
the QCD potential \cite{prd80,prl81},
\begin{gather}
\A_\mu \rightarrow \hA_\mu= A_\mu \n -\frac1g \n \times \pd_\mu \n.
\end{gather}
Notice that the restricted potential is made of two 
parts, the non-toplogical Maxwellian potential  
and the topological Diracian monopole potential. 

\begin{figure}
\includegraphics[height=4.5cm, width=7cm]{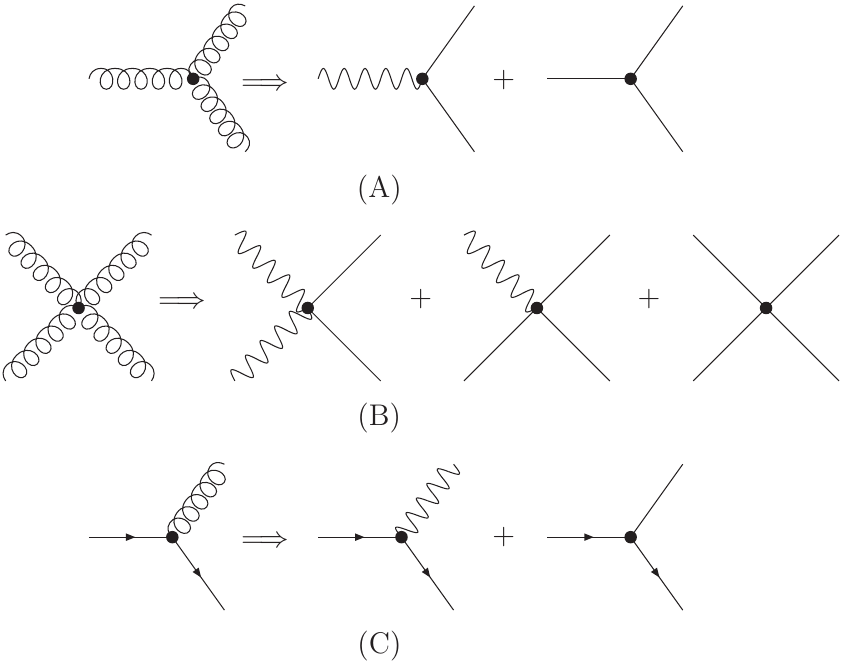}
\caption{\label{3qcdint} The Abelian decomposition of 
the gluon propagators in SU(3) QCD. In (A) and (B) 
the three-point and four-point gluon vertices are decomposed, and in (C) the quark-gluon vertices are decomposed, to 
the nuron and chromon propagators. Here the monopole does 
not appear in the diagrams because it describes 
a topological degree.}
\end{figure}

With this Abelian projection we recover the full 
potential $\A_\mu$ adding the valence potential 
$\X_\mu$ to $\hA_\mu$,
\begin{gather}
\A_\mu=	\hA_\mu +\X_\mu,~~~(\n \cdot \X_\mu =0).
\end{gather}
This is the Abelian decomposition of $\A_\mu$. Notice 
that $\hA_\mu$ retains the full non-Abelian gauge 
degrees of freedom and $\X_\mu$ transforms covariantly 
under the gauge transformation. The Abelian decomposition 
of the QCD gauge potential is graphically shown in 
Fig. \ref{cdec}.

The advantage of the Abelian decomposition is that it 
is gauge independent. Once the Abelian direction is 
chosen, it follows automatically. And we can choose 
any direction to be Abelian. The Abelian decomposition 
of the SU(3) QCD is a bit more compilcated, but straightforward. Here we simply show the Abelian 
decomposition of the gluon propagators of the SU(3) QCD graphically in Fig. \ref{3qcdint}.

The Abelian decomposition allows us to do many things 
that we cannot do in the conventional QCD. First, 
in the pertubative regime it allows us to decompose 
the QCD Feynman diagrams in such a way that color charge consevation is explicit \cite{prd13,epjc19}. This is 
evident in Fig. \ref{3qcdint}. Second, it replaces 
the quark and gluon model of the hadron to the quark 
and chromon model \cite{prd15,prd18}. This is because 
the nuron plays the role of the binding gluon and 
the chromon plays the role of the constituent gluon. 
This clarifies the hadron spectroscopy, in particular 
the glueball spectrum, greatly.  

\begin{figure}
\psfig{figure=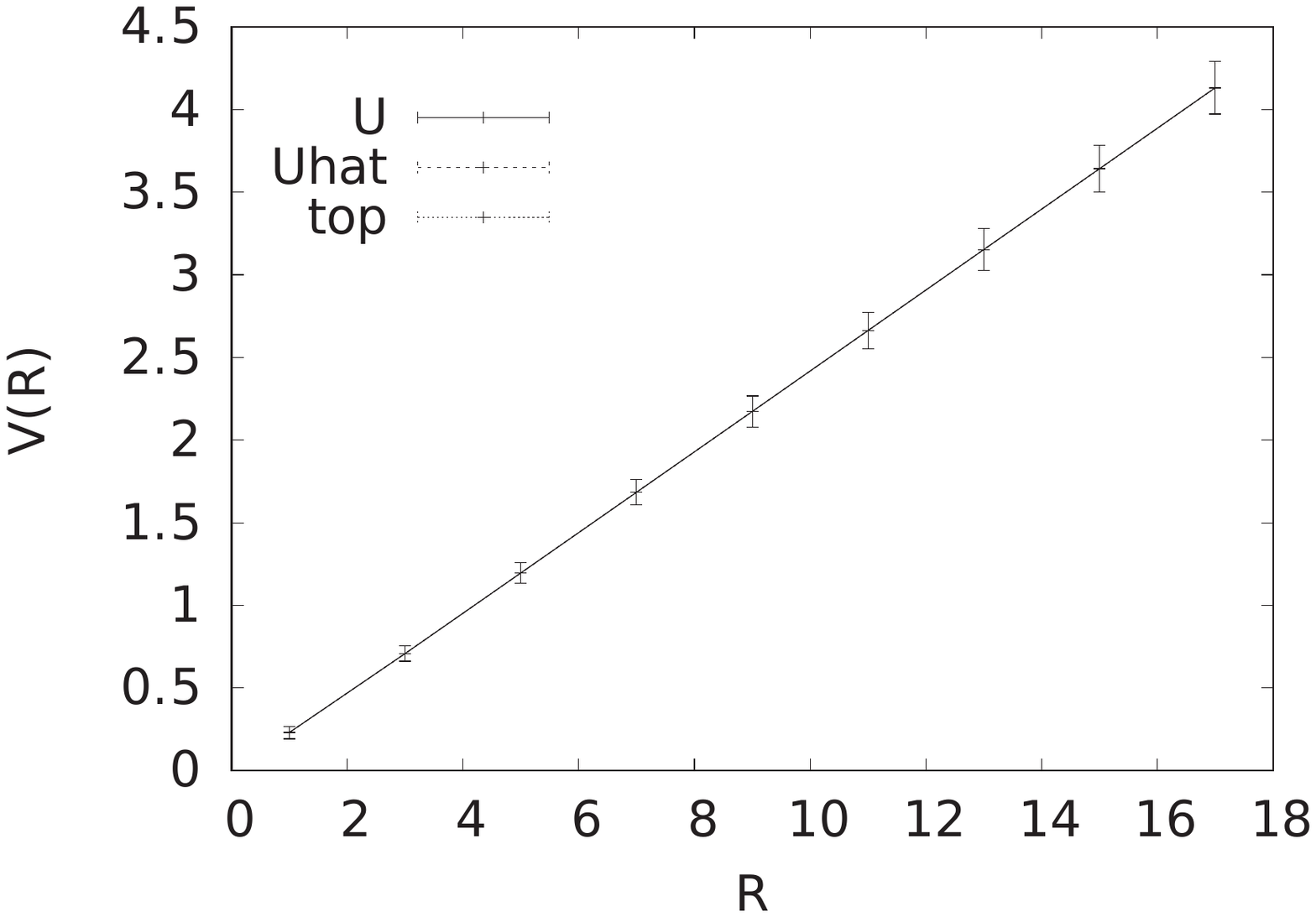, height=5cm, width=8cm}
\caption{\label{cundy} The SU(3) lattice QCD calculation 
which establishes the monopole dominance in the confining force in Wilson loop. Here the confining forces shown 
in full, dashed, and dotted lines are obtained with 
the full potential, the Abelian potential, and 
the monopole potential, respectively.}
\end{figure}

Third, it tells that QCD has a non-trivial Abelian core, 
the restricted QCD (RCD) made of the restricted potential which describes the Abelian subdynamics of QCD but retains 
the full color gauge symmetry. This allows us to prove 
the Abelian dominance, that RCD is responsible for 
the confinement. Using the Abelian decomposition we can calculate the Wilson loop integral and show regourously 
that the restricted potential is responsible for the linear confing force \cite{prd00}. 

Moreover, we can back up this with the lattice QCD calculation. Implementing the Abelian decomposition 
on the lattice, we can calculatethe Wilson loop 
integral numerically with the full potential, 
the Abelian potential, and the monopole potential 
separately, and show that they all generate the linear confining force \cite{cundy,kondo}. This tells that we 
only need the monopole potential to generate the confining force. The result of the lattice calculation is shown 
in Fig. \ref{cundy}. This confirms that the monopole 
is responsible for the color confinement. 

\begin{figure}
\psfig{figure=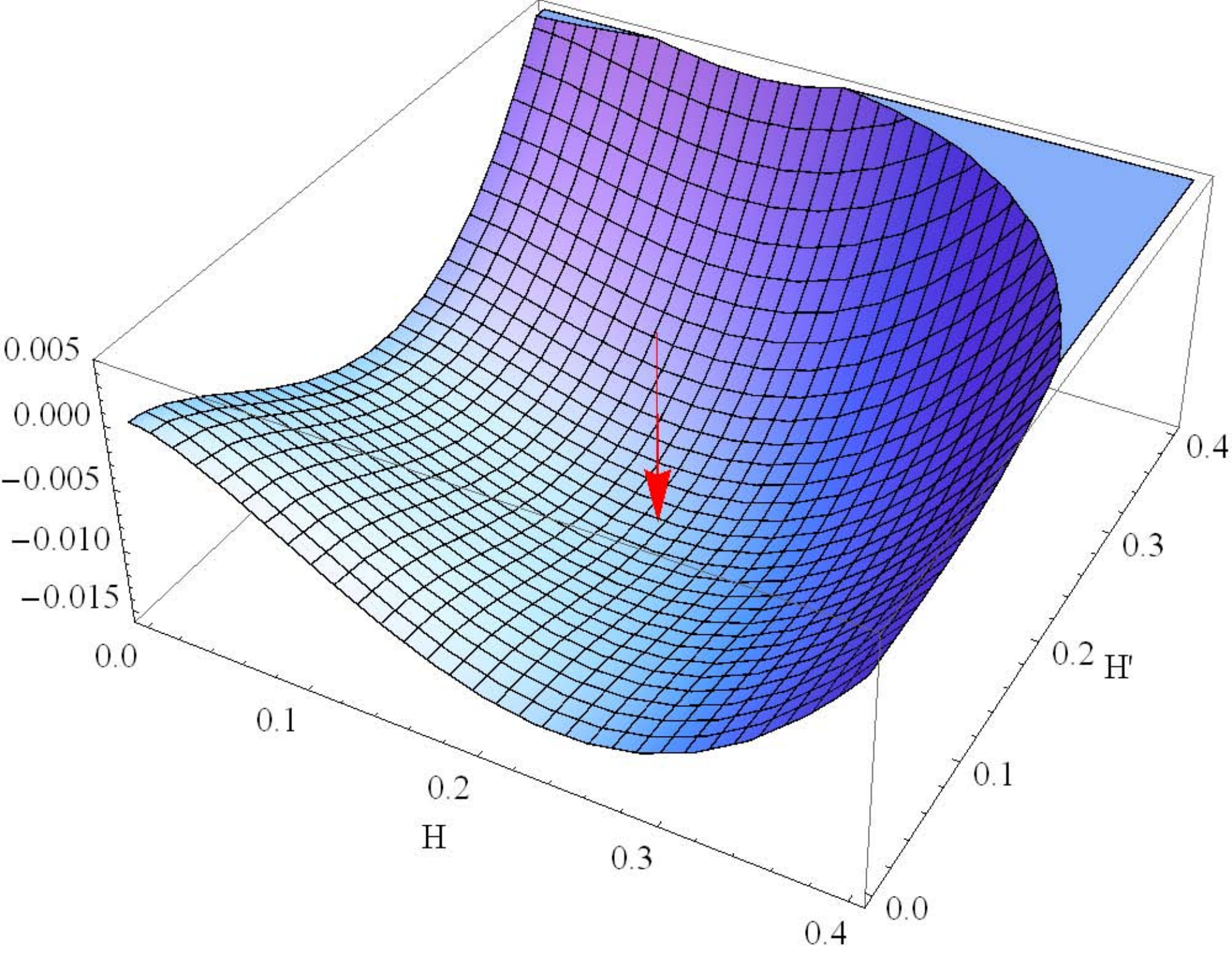, height=4.5cm, width=6.5 cm}
\caption{\label{3pot1} The one-loop SU(3) QCD effective potential which shows that the true minimum of the potential is given by the Weyl symmetric monopole condensation.}
\end{figure}

Furthermore, it puts QCD to the background field formalism, because we can treat the restricted potential as the slow varying classical field and the valence potential as 
the fluctuating quantum field \cite{prd01}. This provides 
us a natrual platform for us to calculate the QCD effective potential and demonstrate the monopole condensation. 
Treating RCD as the classical background and integrating 
out the valence potential under the monopole background, 
we can calculate the one-loop QCD effective potential and prove that the physical vacuum of QCD is given by the stable monopole condensation \cite{prd13,epjc19}. The effective potential of the SU(3) QCD is shown in Fig. \ref{3pot1}. 
This tells that in the perturbative regime (after 
the monopole condensation) QCD is described by two types 
of gluons, the nuron amd chromon.

But, of course, the most important point of the Abelian decomposition is that QCD in the perturbative regime has 
two types of gluons, the color neutral nurons and the colored chromons, which behave differently. If this is true, we must 
be able to show the existence of two types of gluons experimentally. 

\section{Nuron Jet and Chromon Jet}

The prediction of gluon in QCD and the experimental confirmation of the gluon jet was a milestone that justified QCD as the theory of the strong interaction \cite{ellis,gjet}. This implies that, if there really exist two types of 
gluons, QCD must have two types of gluon jets. 

Experimentally the gluon jet tagging has been a very 
complicated business which is not completely understood 
yet \cite{je1,je2,je3,jt1,jt2,jt3,jt4}. The success rate 
of the gluon jet tagging has been only about 70\%. Moreover, the existing gluon jet data are known to have ``anormalies" and ``puzzles" which could not be explained within 
the framework of the conventional QCD \cite{aleph,delphi,hic}. In the following we argue that these anormalies and puzzles 
in the gluon jet experiments could be explained by 
the existence of two types of gluons. In particular, we provide five circumferential evidences which strongly imply the existence of two types of gluons from the existing ALEPH, DELPHI, and CMS gluon jet data. 

To confirm the existence of two types of gluon jets experimentally, we have to know the basic features of 
the nuron and chromon jets. Since two of the eight 
gluons are nurons, one quarter of the gluon jets 
should be the nuron jets and three quarters of 
them should be the chromon jets. If so, how can we 
test the existence of two types of gluon jets
experimentally? 

\begin{figure}
\psfig{figure=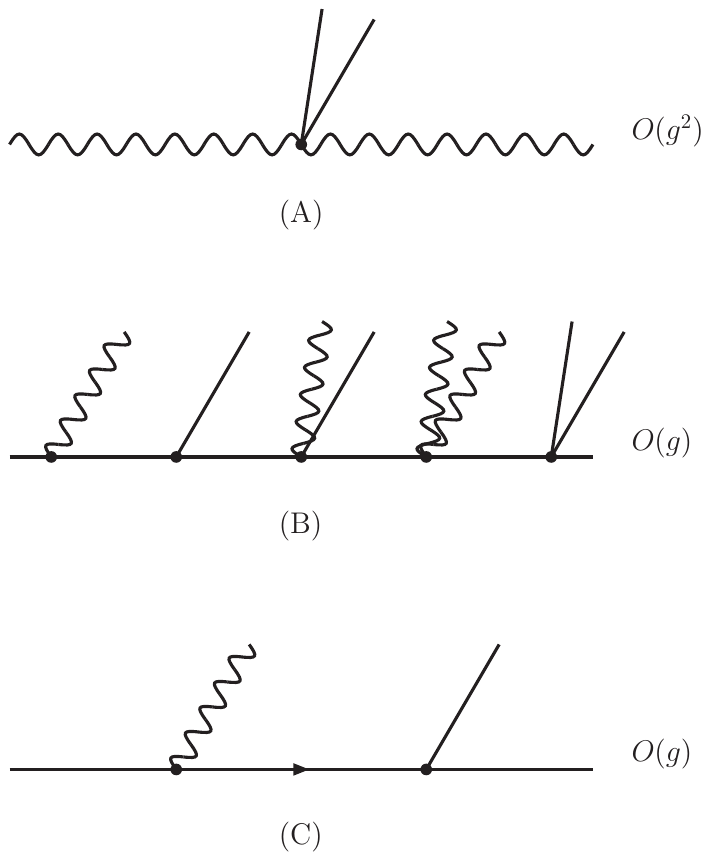, height=5cm, width=7.5cm}
\caption{\label{jet} The parton shower (the soft gluon 
radiation) of hard partons. The strength of the parton 
shower of the nuron jet shown in (A) is of $O(g^2)$, 
but the chromon jet and the quark jet shown in (B) 
and (C) are of $O(g)$. Moreover, the nuron jet has 
only one type of the parton shower, while the chromon 
and quark jets have five and two.}
\end{figure}

To answer this remember that the gluons and quarks 
emitted in the p-p collisions evolve into hadron jets 
in two steps, the soft gluon radiation of the hard 
partons described by the perturbative process and 
the hadronization described by the non-perturbative 
process. Since the hadronization in the second step is basically similar in all jets, the main differences come 
from the parton shower (the soft gluon radiation) in 
the first step. From the Abelian decomposition of 
the Feynman diagram shown in Fig. \ref{3qcdint} we can 
figure out the difference of the soft gluon radiation of 
the nuron, chromon, and quark jets. This is shown 
in Fig. \ref{jet}. 

Clearly the nuron jet shown in (A) has only one type
of soft gluon radiation (i.e., the two-chromon radiation). 
In contrast, the chromon jet in (B) has five different 
types of soft gluon radiations, and the quark jet in (C) 
has two. Moreover, the leading order of the soft gluon radiation of the nuron jet shown in (A) is of $O(g^2)$, 
while that of the chromon and quark jets shown in (B) 
and (C) are of the order of $O(g)$. 

This tells that the nuron jet is expected to behave 
more like the photon jet, sharper than the chromon 
and quark jets with relatively smaller jet radius. 
This is because the nuron has only one soft gluon 
radiation, while the chromon has four more which could 
only broaden the jet. Moreover, this also strongly 
implies that the nuron jet should have considerably 
less particle multiplicity than the chromon and quark 
jets. This must be clear from Fig. \ref{jet}.  

The soft gluon radiation of the nuron jet shown in (A) 
has another important implication. Since it has only one 
type of parton shower made of chromon-antichromon pair, 
it could most likely produce glueball states (more precisely the chromoball states) made of the chromon-antichromon pair, if the energy of the jet becomes close to the mass of 
the chromoball states \cite{prd15,prd18}. On the other hand, the chromon and quark jets shown in (B) and (C) imply 
that they have little chance to produce such states. 
This implies that (when the jet energy becomes close to 
the glueball mass) the nuron jets could produce more 
neutral particles than the chromon and/or quark jets, 
because the chromoballs produce more neutral particles \cite{prd15,prd18}. This provides another characteristic feature of the nuron jet. As we will see, this point 
will have an important consequence in the following.  

\begin{figure}
\psfig{figure=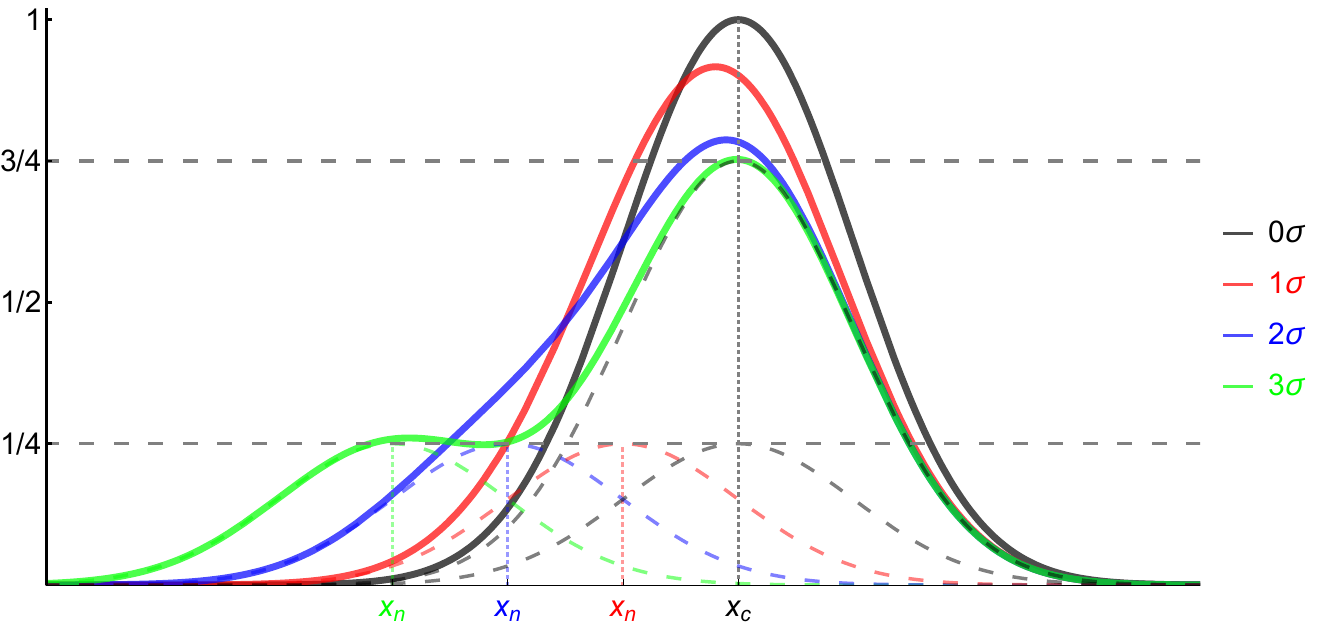, height=5cm, width=8cm}
\caption{\label{gjd0} The expected gluon jet distribution against the particle multiplicity. Here we have plotted 
the overall particle multiplicity when the distance 
between the chromon and nuron peaks becomes one, two, 
and three standard deviations. Notice that the black 
curve represents the well known particle multiplicity 
we have in the conventional QCD.}
\end{figure}

Suppose the nuron jet has less particle multiplicity.
In this case there should be one more local peak on 
the side of the less number of particles in the particle distribution in the gluon jet. So, the particle multiplicity of the gluon jets must have an asymmetry against the main peak axis toward the less number of particles \cite{prd23}. The expected particle multiplicity distribution of the gluon jets is shown in Fig. \ref{gjd0}. In this figure we have assumed (for simplicity) that 
the particle distribution of nuron and chromon jets 
are Gaussian, and plotted the expected particle distribution when the nuron peak is one, two, and three standard deviations away from the chromon peak. Notice 
that the shape of the overall particle distribution of 
the gluon jets crucially depends on the distance between the two peaks. But clearly the figure shows that we have more particles in the left side of the peak. 

For example, when the nuron peak is three standard 
deviation away, the left part of the peak axis has 
64 \% more particles than the right part. And we have 
42 \% more events when the nuron peak is two standard 
deviation away. But when the distance becomes one 
standard deviation, the left part has only 6 \% more 
particles. So the nuron jet could easily be left 
unnoticed when the distance between the nuron and 
chromon peaks becomes less than two standard deviation. 

This should be contrasted with the black curve, which 
represents the particle distribution of the gluon jets 
when the nuron and chromon jets have the same particle multiplicity. This, of course, is the well known particle distribution of the gluon jets we expect in the conventional QCD (without the Abelian decomposition). In this case 
the asymmetry disappears completely. This tells that 
the existence of the reflection asymmetry (tilt) of 
the particle distribution against the peak axis, if exists, can be viewed as a strong indication that there could 
be two types of gluon jets. 

\begin{figure}
\psfig{figure=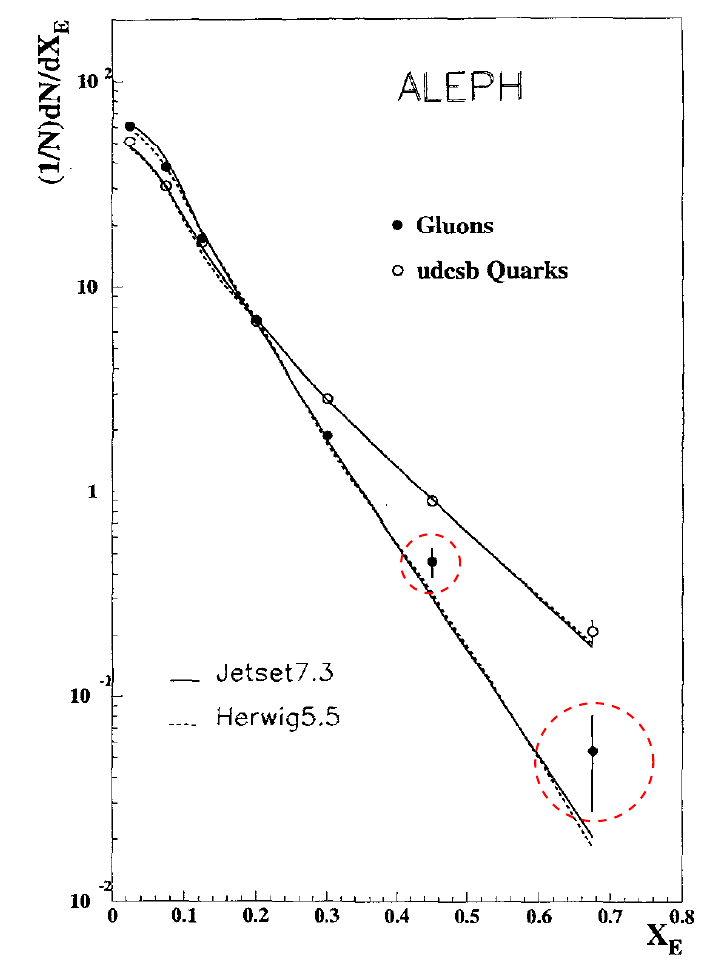, height=5.5cm, width=7cm}
\caption{\label{alephge} The ALEPH data on the energy 
fragmantation of particles of the gluon and quark jets. 
The Monte Carlo predictions of Jetset and Herwig are 
shown in solid and dotted lines, and the experimental 
data of the gluon and quark jets are shown in square 
and solid points. Clearly the data shows an unmistakable excess of particles with largy energy fragmatation 
of gluon jets.}
\end{figure}

There is another way to confirm the existence of 
the nuron jets. Notice that the equipartition of 
energy tells that nuron jets and chromon jets must 
have the same energy in average. So, if the nuron 
jets have less particle multiplicity as we expect, 
the particles coming from the nuron jets must have 
more energy in geneal than those coming from the chromon 
jets. If so, checking the energy distribution of 
the particles coming from the gluon jets and verifying 
the existence of the excess of particles with higher energy, we can show the existence of the nuron jets. 

This means that, without trying to prove the existence 
of the nuron jet and the chromon jet directly, we could tell the existence of two types of gluon jets, checking 
the particle multiplicity and the energy distribution 
of the particles from the existing gluon jets data \cite{uni19,prd23}. This is really remarkable.

\section{Experimental Evidence of Two Types of Gluons: Re-interpretation of ALEPH, DELPHI, and CMS Gluon Jet Data}

The above discussion tells that one quarter of 
the existing gluon jets should actually be the nuron jets which have different jet shape (the sphericity), particle multiplicity, and color dipole pattern. Is there any evidence for this in the existing gluon jet data? The answer is definitely yes. 

\begin{figure}
\psfig{figure=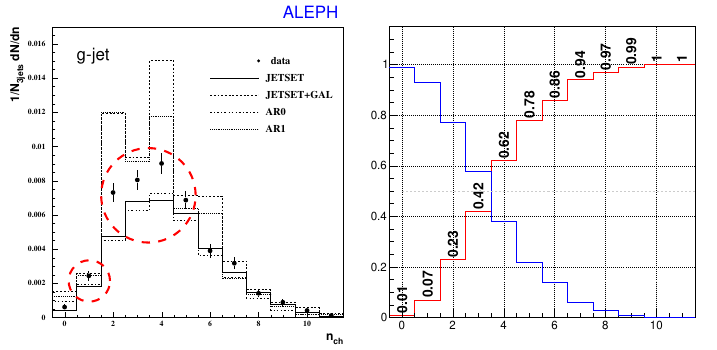, height=5.5cm, width=8cm}
\caption{\label{alephncint} The leftside is the ALEPH 
data on the charged particle multiplicity of the gluon 
jets. The Monte Carlo prediction of Jetset is shown in solid line and the experimental data of the gluon jets 
are shown in solid points. The rightside is the integral 
of the events of the ALEPH data, which shows an asymmetry against the peak axis at $n_c=4$.}
\end{figure}

A best place to look for such evidence is the ALEPH gluon 
jet data coming from the $Z$ boson decay \cite{aleph}
\begin{gather}
e \bar e \rightarrow Z \rightarrow b \bar b g.
\label{qgjet1}
\end{gather}
The ALEPH collaboration has the published results 
which indicate the existence of two types of 
gluons \cite{aleph}, the ``anormaly" in the energy fragmantation of the particles shwn in Fig. \ref{alephge} 
and the ``abnormal" charged particle multiplicity of 
the gluon jets shown in Fig. \ref{alephncint}. Consider 
the energy fragmantation of the particles first. Clearly 
Fig. \ref{alephge} shows that the Monte Carlo predictions 
of the energy distribution of the particles coming from 
the quark jet are in good agreement with the experimental data, but those of the gluon jets have an obvioius disagreement with the experiment, the unexpected excess 
of the more energetic particles shown in red circles. 
This is the well known anormaly of the energy 
fragmantation of the particles in the gluon jets. 

But as we have pointed out, this anormaly is in fact 
exactly what we would expect if we have two types 
of gluons. This is because the particles from 
the nuron jets should have more energy than those 
from the chromon jets, since the nuron jets produces 
less particles. This is precisely what we observe in 
the ALEPH data in Fig. \ref{alephge}. This strongly 
implies that the anormaly is not an anormaly but 
an evidence of the existence of the nuron jet. 

The same ALEPH data provides us another evidence of 
two types of gluon jets, on the (charged) particle multiplicity shown in the leftside of 
Fig. \ref{alephncint}. According to our prediction 
this ALEPH data must have the reflection asymmetry 
against the peak axis, the excess of the less particles, predicted in Fig. \ref{gjd0}. Remarkably, the ALEPH 
data does show this asymmetry. It tells that the peak 
is located at $n_C=4$, but clearly shows the excess 
of the events against the Monte Carlo prediction 
when $n_c$ becomes less than 4. 

\begin{figure}
\psfig{figure=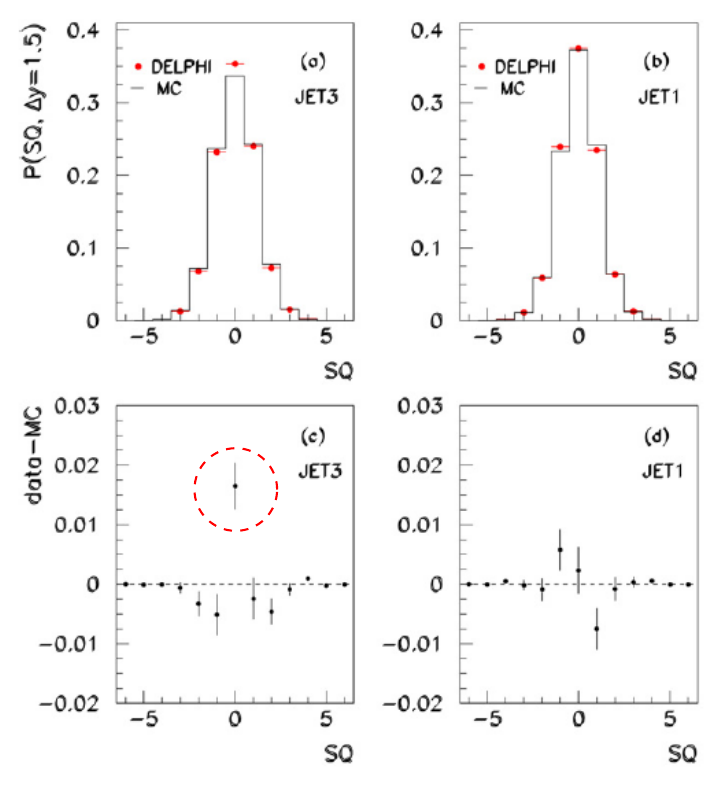, height=6cm, width=8cm}
\caption{\label{delphin} The DELPHI experiment
which shows a clear excess of the neutral particles 
in gluon jet in (a) and (c), which is absent in 
the quark jet in (b) and (d).}
\end{figure}

Moreover, the integral of the experimental data shown 
in the rightside of the figure tells that the particle multiplicity is asymmetric (tilted leftward) against 
the peak axis. It shows that 42 \% of the total number 
of the charged particles are less than 4 but 22 \% are 
more than 4. This tells that the leftside of the peak 
has 20\% more jets. In fact, according to our analysis 
shown in Fig. \ref{gjd0}, the asymmetry implies that 
the experimental data could be made of two peaks, 
the chromon peak with 75 \% population near $n_C=4$ 
and the nuron peak with 25 \% population located 
leftside roughly 1.5 standard deviation away from 
the chromon peak. 

Suppose ALEPH discovered this nuron peak without 
knowing the existence of the nuron jet. In this case 
they would have called this an ``anormaly", because
they would have no way to explain this. So we could 
say that the old ALEPH gluon data actually have two 
anormalies, the anormaly in energy fragmantation 
and the anormaly in particle multiplicity. But this 
second anormaly (in particle multiplicity) has been 
completely unnoticed till now. And this new anormaly 
in the particle multiplicity of the ALEPH gluon jets 
could be viewed as another evidence of the nuron jet. 

So far we have discussed the gluon jet anormalies 
of the ALEPH data. But there exists another type 
of the gluon jet anormaly coming from the DELPHI
experiment \cite{delphi}. Based on the same $Z$ boson 
decay shown in (\ref{qgjet1}), the DELPHI group 
separated the quark jets with purity about 90 \% 
and the gluon enriched jets with purity about 70 \%, 
and observed a clear excess of the neutral particles 
in the gluon jets at low invariant masses less or 
equal to 2 GeV. This is shown in Fig. \ref{delphin}. 
Obviously this is an anormaly. The DELPHI group has interpreted this anormaly as ``an indication that 
the gluon jet might have an additional hithertoo 
undetected fragmantation mode via a two-gluon system"
telling that ``this could be an indication of 
a possible production of gluonic states as predicted 
by QCD" \cite{delphi}.

\begin{figure}
\psfig{figure=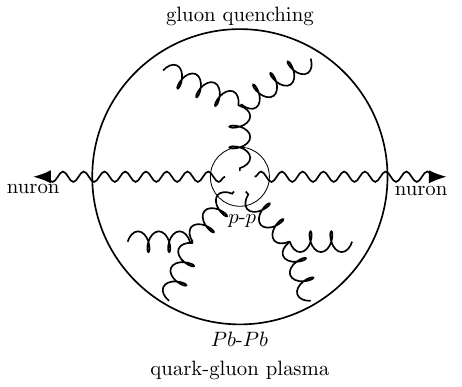, height=6cm, width=7.5cm}
\caption{\label{gquenching} The expected gluon quenching 
in heavy ion collision. The gluons in p-p collision 
can easily escape because the volume of the quark-gluon plazma is small, but in Pb-Pb collision they are 
supposed to be quenched before they escape. Notice,
however, the nurons can easily escape because 
the coupling is weak.}
\end{figure} 

However, this is precisely the characteristic feature 
of the nuron jet that we have mentioned above. 
Obviously the neuron jet shown in (A) of Fig. \ref{jet} tells that (when the jet energy is close to 
the glueball mass) it could most likely produce 
the glueballs (more precisely, the chromoballs) made 
of the chromon-antichromon pair, which produce mostly
the neutral particles. This shows that the figure 
in (A) of Fig. \ref{jet} is not only in line with 
the DELPHI interpretation, but more importantly 
provides a theoretical justification of this 
interpretation. Moreover, this tells that the excess 
of the neutral particles in the gluon jet comes from 
the existence of the neuron jet, which could be 
interpreted as one more supporting evidence of 
the two types of gluon jets. This is remarkable.  

\begin{figure}
\psfig{figure=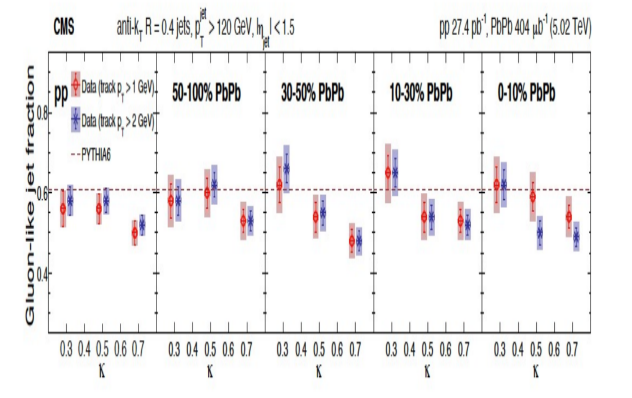, height=6cm, width=8cm}
\caption{\label{cms} The CMS gluon jet data in Pb-Pb 
heavy ion collision compared to the gluon jets in p-p 
collision. The expected gluon quenching in Pb-Pb
collision is hardly detectable.}
\end{figure}

We have another circumferential evidence for two types 
of gluons coming from an unexpected quarter, the CMS 
data on the heavy ion collision \cite{hic}. The gluon 
jets in heavy ion collisions in general are supposed 
to undergo the quenching when they pass through 
the quark-gluon plasma of the heavy ions, so that 
the gluon jets coming out from the heavy ion collisions 
are expected to be less than those coming from the p-p collision. This is the quenching effect schematically 
shown in Fig. \ref{gquenching}. But in the CMS experiment of the Pb-Pb collision, the gluon jets coming from 
the Pb-Pb and p-p collisions have not much difference, which implies that the gluons undergo less quenching 
than expected in heavy ion collision. The CMS result is 
shown in Fig. \ref{cms}. This has been thought to be 
a ``puzzle" \cite{hic}. 

Fortunately, this unexpected puzzle could also be 
explained in the framework of the Abelian decomposition. Since the nurons are color neutral their interaction 
with the quark-gluon plasma must be weaker compared to 
the chromon interaction, as we have explained in 
Fig. \ref{jet}. So the nurons come out almost unquenched 
in the Pb-Pb heavy ion collision. This is shown in 
Fig. \ref{gquenching}. And this could explain 
the puzzle in the CMS heavy ion collision observed 
in Fig. \ref{cms}.    

So far we have discussed four circumferential evidences
for the existence of two types of gluon jets. Each one separately may not be viewed as serious, but together 
they do make a serious case. One could go further and 
check if the ALEPH data has two jet shapes (sharper 
nuron jets and broader chromon jets) and different 
color dipole patterns (ideal color dipole pattern for 
the nuron jets and distorted dipole pattern for 
the chromon jets). 

But in our mind a more important test is to check 
the correlation between the two ALEPH data shown 
in Fig. \ref{alephge} and Fig. \ref{alephncint}. 
Notice that the data which show the anormaly in 
Fig. \ref{alephge} and the asymetry in 
Fig. \ref{alephncint} actually come from the same nuron jets, so that they should be thought as different aspects of the same physics. This means that, if our interpretation is correct, they should 
be correlated. So, confirming this correlation we could increase the credibility of the above interpretation greatly. Unfortunately, ALEPH had no result on this correlation, probably because they 
had no motivation to check this correlation.

\section{Re-analysis of ALEPH Gluon Jet Data by MIT-UoS Group}

The above ALEPF, DELPHI, and CMS gluon jet data stongly indicate the existence of two types of 
gluons. On the other hand, these anormalies have 
not been confirmed by an independent group, so that the data need to be re-analyzed. Recently the MIT-UoS group has reanalyzed the ALEPH data. The purpose was two-fold: Is there really an anormaly in the gluon 
jet data? And if so, does this imply the existance 
of two types of gluons? Now, we discuss the outcome 
of the re-analysis \cite{kcms}.

\begin{figure}
\psfig{figure=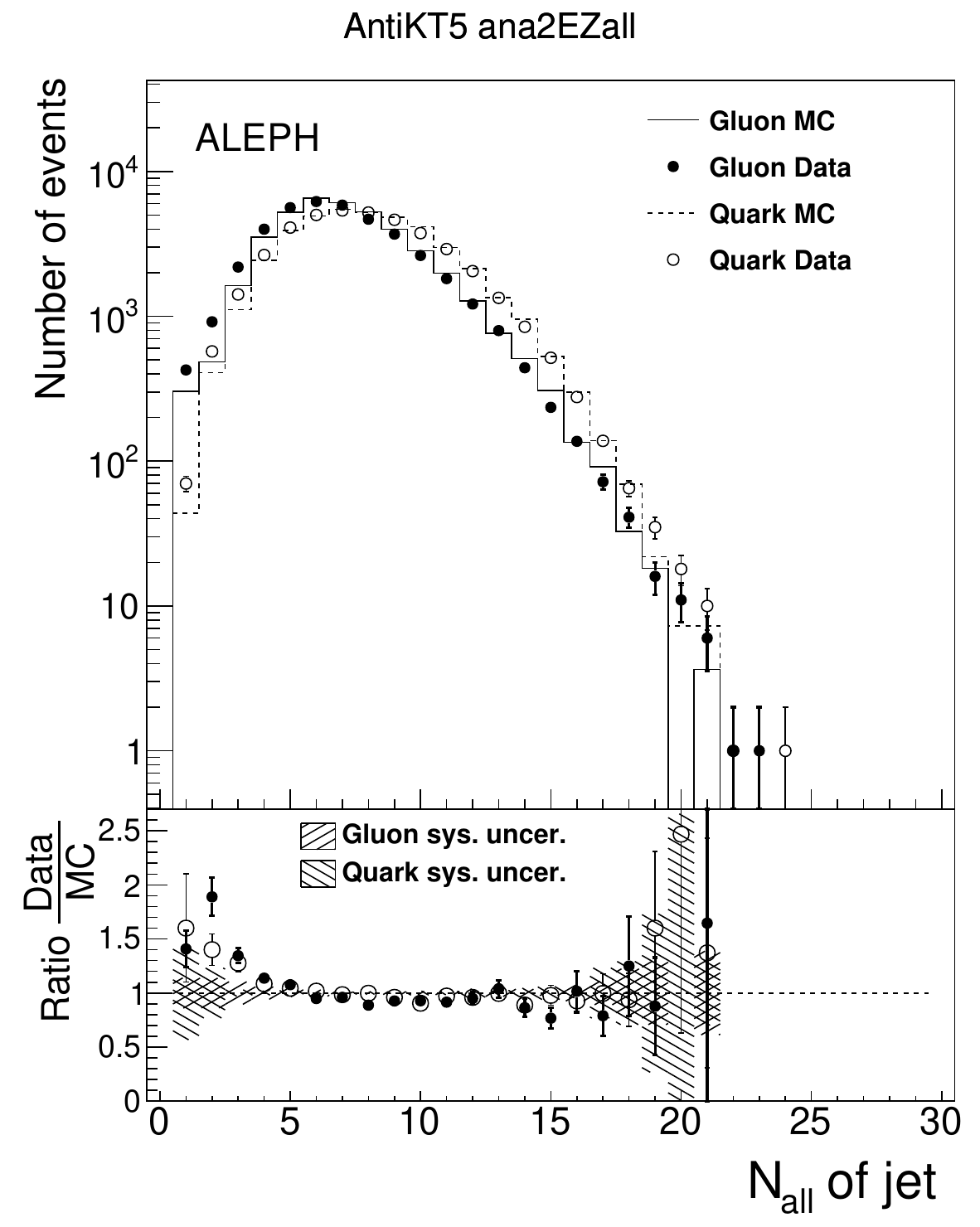, height=6cm, width=8cm}
\caption{\label{knt} The MIT-UoS reanalysis of 
the ALEPH gluon jet data on the total number of the particle multiplicity. The result confirms the excess of the gluon jet events for $n=1,2$, and $3$.}
\end{figure}

From the 2900317 raw ALEPH $b\bar b g$ three jet events coming from the $Z$ boson decay obtained from the years 1991 to 1995, the MIT-UoS group selected 40984 gluon jets using the gluon tagging method similar to the one adopted by the ALEPH. With this they obtained the particle multiplicity distribution of the gluon jets for all, charged, and neutral particles. The result of the particle multiplicity for all particles is shown in Fig. \ref{knt}. 
It has the peak at $N=6$, but also shows that the number 
of the events around $N=2$ and 3 is roughly 2 and 
1.5 times more than the Monte Carlo prediction. This strongly indicates the existence of a second peak around $N\simeq 2.5$. This is precisely the anormaly of 
the particle multiplicity distribution coming from 
the nuron jet that we have discussed. 

They also have re-analyzed the charged and neutral 
particle multiplicity distribution, and confirmed 
the existence of the same trend. For example, for 
the charged particle multiplicity distribution, 
the peak is located at $N=4$, but it has a second peak 
at $N=1$ which indicates the existence of the anormaly 
in the charged particle multiplicity distribution. 
With this we could conclude that the anormaly in 
the particle multiplicity of the ALEPH gluon jet data 
is real, and that we could attribute this anormaly 
to the existence of the nuron jets. It is really 
interesting that the MIT-UoS result is virtually 
identical to our re-interpretation of the published 
result of the ALEPH jet data shown in 
Fig. \ref{alephncint}. 
  
\begin{figure}
\psfig{figure=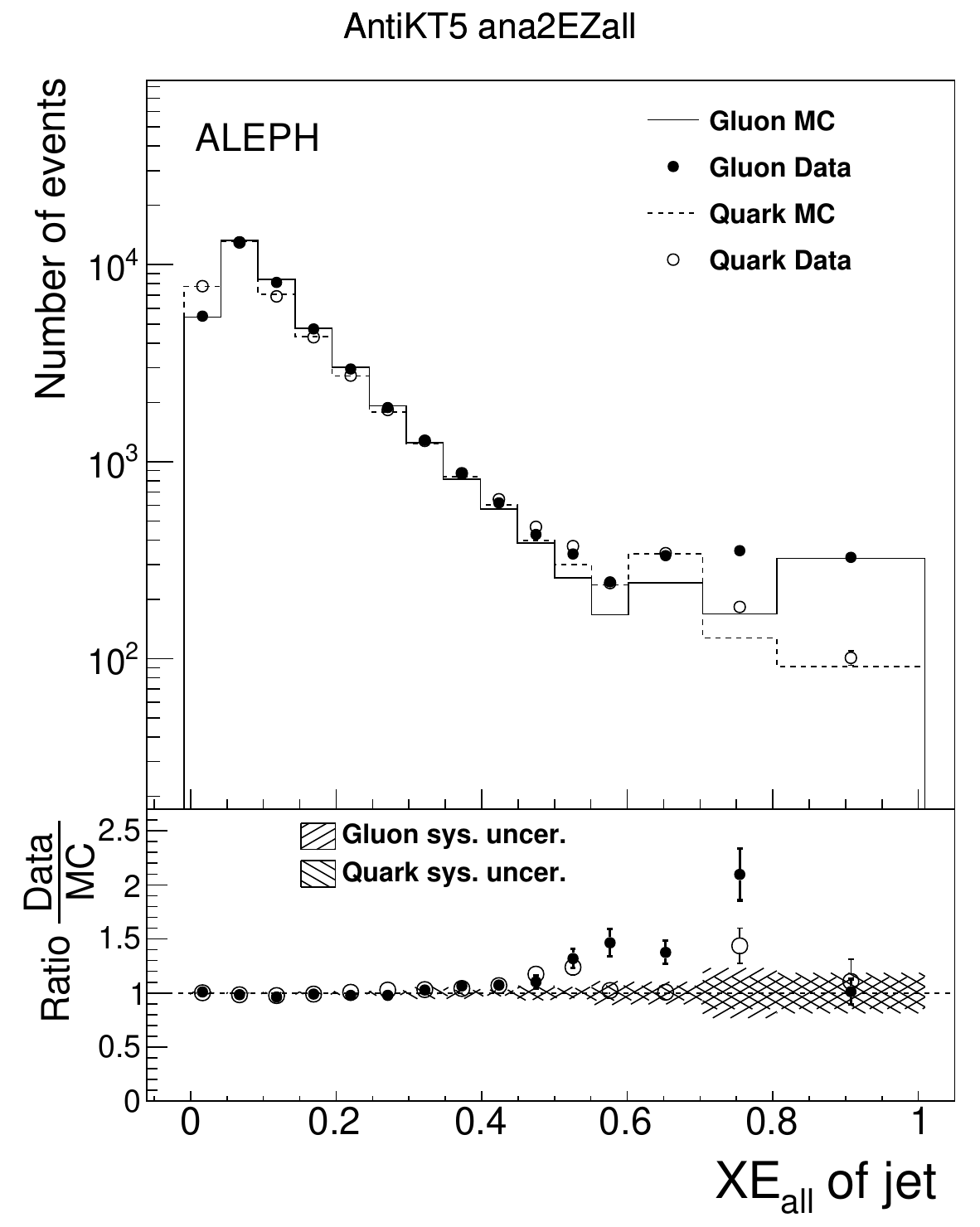, height=6cm, width=8cm}
\caption{\label{kxet} The MIT-UoS re-analysis of the ALEPH gluon jet data on the distribution of the energy fragmantation of the particles. The result confirms the excess of the energy fragmantation $X_E$ above 0.5.}
\end{figure}

With the same ALEPH data, the MIT-UoS group has 
re-analized the energy fragmantation of all, charged, 
and neutral particles coming from the gluon jets. 
The result of the energy fragmantation of all 
particles is shown in Fig. \ref{kxet}, which should 
be compared with the ALEPH result shown in 
Fig. \ref{alephge}. Here again, the result shows 
that the gluon jet energy fragmantation above 0.5 
is 1.5 to 2 times bigger than the Monte Carlo 
prediction, which confirms the existence of the anormaly 
of the ALEPH gluon jets in large $X_E$ above 0.5. 
And obviously, this could be attributed to the existence 
of the nuron jets which have more energetic particles.  

\begin{figure}
\epsfig{file=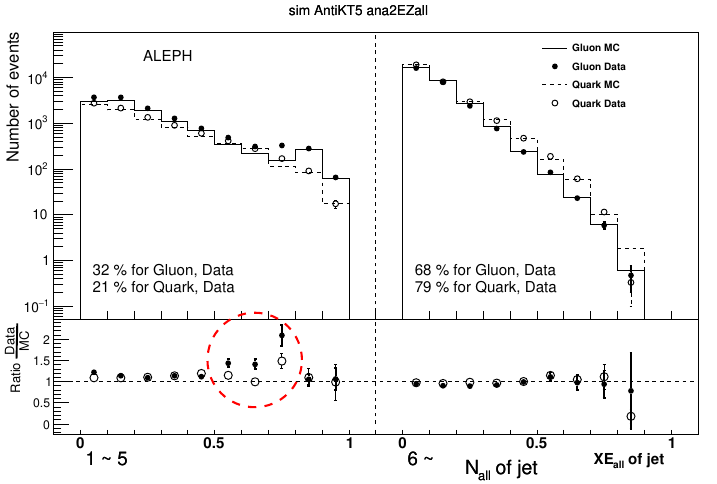,height=6cm,width=8cm}
\caption{\label{conet} The correlation of the gluon jets between the charged particle multiplicity and the energy fragmantation of particles of the ALEPH experiment.}
\end{figure}

If this interpretation is correct, the two anormalies 
should be correlated, because they come from the same 
nuron jets. Remarkably, we can demonstrate this 
correlation from the ALEPH data. The MIT-UoS group 
have checked this correlation between all, charged, 
and neutral particle multiplicity and the corresponding 
energy fragmantation, and confirmed the existence of 
the correlation. The correlation for all particles 
is shown in Fig. \ref{conet}. 

It shows that 32 \% of the events below $N=6$ fully 
expain the anormaly in the large energy fragmantation. Indeed, since this 32 \% also contains the tails of 
the chromon jets which have high particle multiplicity, 
we can tells that the nuron jets which consist of 25 \% 
of the gluon jets with the low particle multiplicity 
less than $N=6$ can fully account for the large energy fragmantation over 0.5 shown in Fig. \ref{kxet}. This confirms that there exists a strong correlation between 
the low particle multiplicity events and high energy fragmantation events. This is precisely the correlation predicted by the existence of the nuron jets. The MIT-UoS group also confirmed the existence of the similar correlation for the charged and neutral particles.  

Clearly this confirmation of the correlation between 
the low particle multiplicity events and high energy fragmantation events is the most compelling evidence 
of the exisence of the nuron jet. Of course, 
the excess of the low particle multiplicity shown in 
Fig. \ref{alephncint} and Fig. \ref{knt} or the excess 
of the high energy fragmantation shown in 
Fig. \ref{alephge} and Fig. \ref{kxet} independently 
implies the existence of the nuron jets. However, 
there is no reason why they should be correlated, 
unless they come from the same origin as we have 
predicted based on the Abelian decomposition of QCD. 
And the above correlation assures that indeed our interpretation is correct. 

\section{Discussion}

The Abelian decomposition of QCD tells that there are two types of gluons, the color neutral nurons and colored chromons, in QCD. In this paper we have studied the four 
gluon jet anormalies and puzzles in the ALEPH, DELPHI, and CMS gluon jet experiments, and argued that these anormalies and puzzles could in fact be interpreted as the evidence of the existence of two types of the gluon jets. Based on the Abelian decomposition of QCD, we have argued that the excess of the low particle multiplicity and high energy fragmantation in the ALEPH data, the excess 
of the neutral particles in the gluon jets in the DELPHI data, and the absence of the gluon quenching in the Pb-Pb collision in the CMS data could be interpreted as the circumferential evidence of two types of gluons in QCD. 

More importantly, we have discussed the existence of a new correlation between the low particle multiplicity events and high energy fragmantation events in the ALEPH data, which is obtained by the MIT-UoS group re-analyzing the ALEPH gluon jet data \cite{kcms}. This correlation can be viewed a most important 
and persuasive evidence of the existence of the nuron jet. This is because this shows that both excesses come from the same origin, 
the nuron jet. 

Admittedly, the above evidences may not be viewed 
as a direct evidence of the existence of two types 
of gluons, since they do not identify the nuron and 
chromon jets separately. To have the direct evidence, 
we need new Monte Carlo simulations, new Pythia and 
FastJet predictions based on the Abelian decomposition 
of QCD, and confirm the existence of two types of gluon 
jets. This is beyond the scope of this paper. 
Nevertheless the above circumferential evidences 
strongly implies that indeed there could be two types 
of gluons in QCD. This provides us an irresistable 
motivation to re-analyse all existing gluon jet data 
more carefully. 

There are different ways to test the existence of 
two types of gluon jets. One is the jet shape. As we 
have pointed out, the nuron jet is expected to have 
a sharper jet shape, while the chromon jet has 
broader jet shape. So we could test the existence 
of the nuron jet studying the jet shape of the gluon 
jets. Another important feature of the nuron jet 
is that it has different color flow. Clearly 
the chromons and quarks carry color charge, but 
the nurons are color neutral. So the nuron jet 
must have different color flow. In fact, 
Fig. \ref{jet} suggests that the nuron jet generates 
an ideal color dipole pattern, but the other two 
jets have distorted dipole pattern. But as far as 
we understand, these studies have not been done 
by any experimental group. 

It should be emphasized that, in doing so we do not 
need any new experiment. LHC produces billions of hadron 
jets in a second, and ATLAS and CMS have already filed 
up huge data on gluon jets. Moreover, DESY, LEP, 
and Tevatron have old data on three jet events which 
could be used to confirm the existence of the nuron 
and chromon jets. Here again the simple number 
counting strongly suggests that one quarter of 
the gluon jets coming from the three jets events 
should actually be the nuron jets which do not fit 
to the conventional gluon jet category. 

There have been many experiments on the gluon jets, 
and one might wonder why they could not find two types 
of the gluon jets. There could be many reasons, but 
the most important reason might be that there was no 
theoretical motivation for them to do so, probably 
because they were unaware of the possibility of two 
types of gluons in QCD. 

As we have emphasized, the Abelian decomposition does 
not change QCD, but reveals the hidden structures of 
QCD. It tells that QCD has an Abelian core RCD, and 
QCD can be viewed as RCD which has the chromons as 
colored source \cite{prd80,prl81}. It decomposes 
the QCD Feynman diagrams in such a way that color 
conservation is explicit \cite{prd01}. It allows us 
to prove the Abelian dominance and the monopole 
condensation \cite{prd13,epjc19}. And it replaces 
the quark and gluon model by the quark and chromon 
model \cite{prd15,prd18}.

But most importantly, it tells that there are two 
types of gluons, the color neutral nurons and colored 
chromons which behave differently. A common prejudice 
in the conventional QCD is that the eight gluons are 
supposed to be indistinguishable, and nobody has doubted 
this popular wisdom. But the Abelian decomposition tells 
that this may not be true. In this paper we challenged 
this wisdom and showed how to back up this experimentally 
by discussing five evidences of two types of gluons. 
Our findings in this paper strongly implies that we 
may need a totally new interpretation of QCD.    

At this point one might ask what are the gluon jets 
identified so far by experiments. Most probably they 
are the chromon jets, because the chromon jet has 
the characteristics of the gluon jet in the conventional 
QCD. This is evident from Fig. \ref{jet}. A more 
interesting question is why they have not found the nuron 
jet. Obviously, they have not seriously searched for 
the nuron jet, because they had no motivation to do so. Moreover, the nuron jets consists of only 25 \% of 
the gluon jets, and might have been systematically 
excluded because they do not fit to the typical gluon 
jet.

In fact, the sucess rate of the gluon jet tagging 
have been only around 70 \%, which implies that 
the experiments have about 30\% of uncertainty on 
the gluon jet. We believe that one reason for this 
uncertainty is because 25 \% of the gluon jets are 
actually made of the nuron jets which behave differently. 
So it is really remarkable that the ALEPH, DELPHI, 
and CMS gluon jet data were able to record evidences 
for two types of gluon jets in spite of this uncertainty. 

In this paper we have discussed the experimental verification of the Abelian decomposition. But the physical applications of the Abelian decomposition is as important as the experimental verification. Obviously 
a natural place to apply the Abelian 
decomposition is the parton distribution in QCD and have a new parton distribution function for hadrons, in particular for 
the proton parton distribution function. In this direction there has been an important progress. Recently the theory group of Institute of Modern Physics of Chinese Academy of Science has re-analized 
the Dokshitzer-Gribov-Lipatov-Altarelli-Parisi 
(DGLAP) evolution equation of the gluon parton 
distribution in proton, and suceeded to 
decompose it and obtain two new evolution equations for the nuron and chromon 
saparately \cite{chen}. This could change our 
understanding of the parton distribution and evolution in QCD drastically. For this to happen, of course, we first need to confirm the existence of two types of gluons in QCD experimentally. And this was precisely what 
we tried to do in this paper.

The experimental discovery of the gluon 
jet was a big step for us to understand 
QCD \cite{wil,ellis}. Clearly the confirmation of two types of gluons will be another giant step for QCD. If confirmed, this could be a most important discovery in QCD which could 
extend the horizon of our understanding of nature one step further.  

{\bf Acknowledgements}

~~~The author thanks Yen-Jie Lee for informing the CMS data on the gluon quenching anormaly in heavy ion collision, and Inkyu Park and Youngkwon Jo for sharing the result of the MIT-UoS re-analysis of the ALEPH data and allowing to use some of them in this paper. The work is supported in part by National Research Foundation of Korea funded by the Ministry of Science and Technology (Grant 2022-R1A2C1006999) and by Center for Quantum Spacetime, Sogang University.


\begin{thebibliography}{99}
\bibitem{wil} D. Gross and F. Wilczek, Phys. Rev. Lett. 
{\bf 30}, 1343 (1973); H. Politzer, Phys. Rev. Lett. 
{\bf 30}, 1346 (1973).
\bibitem{ellis} J. Ellis, M. K. Gaillard, and G. G. Ross, Nucl. Phys. {\bf B111}, 253 (1976).
\bibitem{gjet} R. Brandelik et al. (TASSO Collaboration),
Phys. Lett. {\bf B86}, 243 (1979); D. P. Barber et al. 
(MARK-J Collaboration), Phys. Rev. Lett. {\bf 43}, 830 (1979); C. Berger et al. (PLUTO Collaboration), Phys. Lett. {\bf B86}, 418 (1979).	

\bibitem{nambu} Y. Nambu, Phys. Rev. {\bf D10}, 4262 
(1974); S. Mandelstam, Phys. Rep. {\bf 23C}, 245 (1976); 
A. Polyakov, Nucl. Phys. {\bf B120}, 429 (1977).

\bibitem{prd80} Y. M. Cho, Phys. Rev. {\bf D21}, 1080 (1980); Y. S. Duan and M. L. Ge, Sci. Sinica {\bf 11}, 1072 (1979).
\bibitem{prl81} Y. M. Cho, Phys. Rev. Lett. {\bf 46}, 
302 (1981); Phys. Rev. {\bf D23}, 2415 (1981).

\bibitem{thooft} G. 't Hooft, Nucl. Phys. {\bf B190}, 
455 (1981).
\bibitem{prd00} Y. M. Cho, Phys. Rev. {\bf D62}, 
074009 (2000). 

\bibitem{prd01} W. S. Bae, Y. M. Cho, and S. W. Kim, 
Phys. Rev. {\bf D65}, 025005 (2001).

\bibitem{prd13} Y. M. Cho, Franklin H. Cho, and J. H. Yoon, Phys. Rev. {\bf D87}, 085025 (2013).
\bibitem{epjc19} Y. M. Cho and Franklin H. Cho,  
Euro Phys. J. {\bf C79}, 498 (2019).

\bibitem{uni19} Y. M. Cho, Universe {\bf 5}, 62 (2019).
\bibitem{prd23} Y. M. Cho, Pengming Zhang, and Liping Zou, Phys. Rev. {\bf D107}, 054024 (2023).

\bibitem{prd15} Y. M. Cho, X. Y. Pham, Pengming Zhang, 
Ju-Jun Xie, and Li-Ping Zou, Phys. Rev. {\bf D91}, 
114020 (2015).
\bibitem{prd18} Pengming Zhang, Li-Ping Zou, and Y. M. Cho, Phys. Rev. {\bf D98}, 096015 (2018). 

\bibitem{cundy} N. Cundy, Y. M. Cho, W. Lee, 
and J. Leem, Phys. Lett. {\bf B729}, 192 (2014); 
Nucl. Phys. {\bf B895}, 64 (2015).
\bibitem{kondo} S. Kato, K. Kondo, T. Murakami, A. Shibata, T. Shinohara, and S. Ito, Phys. Lett. {\bf B632}, 326 (2006); {\bf B645}, 67 (2007); {\bf B653}, 101 (2007); {\bf B669}, 107 (2008). 

\bibitem{je1} ATLAS Collaboration, Eur. Phys. J. 
{\bf C73}, 2676 (2013); {\bf C74}, 3023 (2014); 
{\bf C75}, 17 (2015). 
\bibitem{je2} CMS Collaboration, Eur. Phys. J. 
{\bf C75}, 66 (2015); Phys. Rev. {\bf D92}, 032008 (2015). 
\bibitem{je3} ATLAS Collaboration, Eur. Phys. J. 
{\bf C76}, 322 (2016); Phys. Rev. {\bf D96}, 072002 (2017).

\bibitem{jt1} J. Gallicchio and M. Schwartz, Phys. Rev. Lett. {\bf 107}, 172001 (2011); A. Larkoski, G. Salam, and J. Thaler, JHEP, {\bf 06}, 108 (2013).
\bibitem{jt2} B. Bhattacherjee, S. Mukhopadhyay, M. Nojiri, Y. Sakaki, and B. Webber, JHEP, {\bf 04}, 131 (2015); D. de Lima, P. Petrov, D. Soper, and M. Spannowsky, Phys. Rev. {\bf D95}, 034001 (2017). 
\bibitem{jt3} E. Metodiev and J. Thaler, Phys. Rev. Lett. {\bf120}, 241602 (2018); J. Davighi and P. Harris, Euro. Phys. J. {\bf C78}, 334 (2018).
\bibitem{jt4} P. Gras et al. JHEP {\bf 07}, 091 (2017);
A. Larkoski and E. Metodiev, JHEP {\bf 10}, 014 (2019);
CMS Collaboration, JHEP {\bf 01}, 188 (2022). 

\bibitem{aleph} D. Buskulic et al. (ALEPH Collaboration), Phys. Lett. {\bf B384}, 353 (1996).
\bibitem{delphi} J. Abdallah et al. (DELPHI Collaboration), Phys. Lett. {\bf B643}, 147 (2006).

\bibitem{hic} CMS Collaboration, JHEP {\bf 07}, 115 (2020).

\bibitem{kcms} Yen-Jie Lee, Youngkwon Jo, Ingyu Park, and Yi Chen, arXiv;2501.00000 [hep-ph], JHEP, to be published. 

\bibitem{chen} Y. Yang, W. Kou, X. Wang, Y. Cai, 
and X. Chen, Euro. Phys. J. {\bf C84}, 924 (2024).


\end{thebibliography}
\end{document}